\documentclass[conference]{IEEEtran}
\IEEEoverridecommandlockouts
\usepackage{hyperref}
\usepackage{cite}
\usepackage{amsmath,amssymb,amsfonts, amsthm}
\usepackage{algorithmic}
\usepackage{graphicx}
\usepackage{caption}
\usepackage{subcaption}
\usepackage{textcomp}
\usepackage{xcolor}
\usepackage{url}
\usepackage{tikz}
\usepackage{svg}

\usetikzlibrary{positioning, arrows.meta, calc, backgrounds}
\usepackage{float}

\def\BibTeX{{\rm B\kern-.05em{\sc i\kern-.025em b}\kern-.08em
    T\kern-.1667em\lower.7ex\hbox{E}\kern-.125emX}}

\newcommand{\Reconst}{\mathsf{Reconst}}

\newcommand{\equalcontrib}{\textsuperscript{*}}

\theoremstyle{definition}

\theoremstyle{remark}

\theoremstyle{plain} 

\begin{document}

\title{Efficient Data Availability Sampling via Coded Distributed Arrays
}

 \author{
  \IEEEauthorblockN{Dang Pham Minh\equalcontrib\thanks{\equalcontrib Both authors contributed equally to this work.}}
 \IEEEauthorblockA{\textit{Independent Researcher} \\
 Hanoi, Vietnam}
 \and
 \IEEEauthorblockN{Hung Vuong Huu\equalcontrib}
 \IEEEauthorblockA{\textit{Independent Researcher} \\
 Hanoi, Vietnam}
 \and
 \IEEEauthorblockN{Duc A. Tran }
 \IEEEauthorblockA{\textit{University of Massachusetts} \\
 Boston, USA}
 }

\maketitle

\thispagestyle{plain} 

\setcounter{page}{1} 

\pagestyle{plain}

\begin{abstract}
Data availability is a fundamental bottleneck in modern blockchain networks. Most blockchain systems  rely on a full-replication model, which requires downloading of a full block to verify its availability. This model does not scale with block size because  every node must handle large volumes of data, leading to slower block propagation, duplicated data transfer, and longer consensus agreement. This issue is well-known in Ethereum, where layer-2 rollups publish  data directly into the chain. To overcome, Ethereum adopts Data Availability Sampling (DAS) to let nodes keep only a small fragment of the data while still ensuring availability. Prior work on DAS has focused   on   cryptographic foundations. Meanwhile, the peer-to-peer network layer that provides Byzantine-tolerant and scalable mechanisms for discovery and routing of DAS fragments is  underexplored.  We propose CDA, a new design for DAS based on coded distributed arrays that leverages network coding to ensure both robustness and efficiency. Our evaluation study compares CDA to RDA, the latest DAS development of Ethereum, showing an improvement of several times better. 
\end{abstract}


\section{Introduction}
Blockchain networks  face the trilemma of scalability, security, and decentralization. In Ethereum, scalability is primarily pursued through Layer-2 (L2) rollup systems.
In a typical rollup design, a sequencer aggregates off-chain transactions, computes the updated rollup state, and publishes the data required for state verification to the base chain.
This data is only needed for a limited time window, during which nodes can verify state transitions and generate validity proofs.
Yet this data is stored permanently as part of block data, imposing long-term storage and processing overhead on full nodes despite ephemeral relevance.

EIP-4844 \cite{eip4844} introduces a solution for this by defining \emph{blobs}. Large data objects are processed exclusively by the Consensus Layer, which handles block propagation and consensus, rather than the Execution Layer. Blobs exist only for a bounded retention window sufficiently long for L2 verification; after that they may be  discarded. The objective is to increase the number of blobs that can fit into each block, enabling 
higher L2 throughput. However,  if every node in the network were required to store all blobs in full, increasing blob count would quickly become infeasible.

As full data replication cannot scale, Ethereum recently (and other blockchain networks follow similarly)  adopts the method of \emph{Data Availability Sampling} (DAS). 
DAS allows nodes to keep only small fragments of blob data such that availability can be verified 
probabilistically by sampling random positions and cross checking them with peers. Specifically, to publish a blob,
the block proposer (or an external builder acting on its behalf) erasure encodes the blob data  and disseminates the resulting fragments to the network. A verifier node only needs to successfully retrieve sufficient  fragments in order to conclude  that the entire blob is reconstructible.

To realize DAS effectively,   challenges are due to  not only cryptographic guarantees but also how to make   block production and data propagation fast and efficient.
Delays in fragment dissemination result in longer confirmation finality  and weaker availability guarantees.
If availability is not established within time limits, rollup systems cannot reliably validate state transitions and generate fraud proofs.

\textbf{Today's limitations. } Existing DAS developments on the cryptographic layer  are usually built upon erasure codes, polynomial commitments, and sampling guarantees \cite{hasw2023, frida2024, peerdas2024, zoda2025, accidental2025}. They operate under abstract models assuming that sampling queries are resolvable without considering  factors in dissemination delay, retrieval latency, and network dynamics.
To improve  block production latency, one way is to reconstruct the block building workflow \cite{pandas2025} via  centralized or semi-centralized builders. This approach  does not consider the network-layer aspects of  P2P dissemination or sampling protocols.

Most recently, in 2025, a Byzantine-tolerant and sampling-efficient P2P layer for DAS based on the concept of robust distributed arrays (RDA) was introduced by Ethereum Foundation (Feist et al. \cite{rda_paper}). The idea is 1) to organize the nodes as a grid, 2) divide the data block into chunks, each mapped to a grid column, and 3) replicate each chunk in all nodes in the assigned column. RDA offers excellent sampling efficiency and provable Byzantine security. However, the full-column replication requirement is too strong, incurring  excessive storage and communication costs. 

\textbf{Contributions. } Our main problem is to design a DAS solution that jointly optimizes 
storage efficiency, communication overhead, Byzantine robustness, and end-to-end DAS correctness. This  is a gap in the literature.
We propose a  solution called 
CDA (shorthand for Coded Distributed Array), which also organizes nodes into a P2P grid as RDA but differentiates in how a block is coded, published, and verified. CDA is scalable, provably secure, and compatible with existing builder-proposer workflows and erasure-coded blob structures. 
   Specifically, we make the following contributions:
\begin{itemize}
    \item  Rather than fully replicate an entire chunk across a column, we   encode the data using Random Linear Network Coding (RLNC) \cite{multicast_rlnc} such that only coded-pieces of the chunk are disseminated to the column. This results in multi-time improvements in both storage and dissemination overheads.   
    \item The way that a chunk is encoded using RLNC needs a new commitment scheme. We propose one based on KZG offering small commitment size and efficient verification of data validity. Combining this with a robust structure for the P2P network layer, our DAS system is provably secure against Byzantine attacks.
    \item An evaluation study is conducted through realistic benchmarks and simulations. Numerical results confirms our theoretical properties and demonstrate CDA's multi-time superiority to RDA. For example, CDA is more than $5\times$ better in storage, $2\times$ in dissemination cost, and $1.4\times$ in data-synchronization cost when new nodes join.
\end{itemize}

The remainder of this paper is organized as follows. 
Section~\ref{sec:related} reviews related work. 
Section~\ref{sec:prelim} introduces essential background on DAS and RLNC necessary for our research development. 
Section~\ref{sec:design} presents the details of CDA. Numerical results of evaluation is discussed in  
Section~\ref{sec:evaluation}.  
The paper is concluded in   Section~\ref{sec:conclusion}.

\section{Related Work}
\label{sec:related}

Distributed Hash Tables (DHT), via methods such as Kademlia~\cite{kademlia2002} and Chord~\cite{chord2001}, are often discussed as a plausible substrate for  DAS thanks to their distributed-indexing efficiency.  For example,   Cortés-Goicoechea et al.~\cite{kademlia2024} studied the feasibility of integrating a Kademlia DHT into Ethereum’s DAS.  
Their results indicate that while DHT can support data sampling, disseminating large data structures  incurs substantial latency. Also, DHT assumes honest participation; making it robust against adversaries, including data fabrication and Sybil attacks, remains a difficult challenge due to multi-hop routing and  neighborhood structure. Efforts such as Honeybee~\cite{honeybee2024} aim to mitigate these issues, but still relying on honest-majority assumptions. In contrast,  we seek a design that can work with any small honesty fraction (provided a minimal number of honest nodes are active).

For blockchain networks, the idea of verifying transaction data's availability through sampling of erasure-coded data was first introduced by Al-Bassam et al.~\cite{fraudproofs_da} and later adopted by LazyLedger and evolved into Celestia \cite{lazyledger2019}. Celestia is a data availability layer to enable modular blockchain networks,  ensuring that data published to the chain remains available and cannot be withheld by any validator.  Celestia allows validators to collectively guarantee data availability without requiring every validator to download the full block data. Celestia has two types of nodes, super and light. The set of super nodes serves as a centralized servicer for storing and retrieving complete blob data. Light clients verify data availability by issuing sampling queries to these super nodes.

Subsequently, the Ethereum ecosystem has explored alternative DAS designs that avoid centralized reliance on the super nodes. One such proposal is PeerDAS~\cite{peerdas2024}, which has been integrated into the Fusaka~\cite{fusaka} integration in December 2025. PeerDAS relies on DHT-based retrieval and gossip-based dissemination, assigning columns of the erasure-coded data matrix to dedicated GossipSub~\cite{gossipsub2020} channels. Both mechanisms rely on multi-hop communication, hence incurring high latency for sampling requests and data propagation. This is worsened by the presence of Byzantine participants, leading to failure of availability verification. 

PANDAS \cite{pbs} was proposed to  address this multi-hop inefficiency of PeerDAS. In PANDAS, the responsibility of block dissemination is delegated to block builders (not block proposers) and instead of gossips, the dissemination is direct from a block builder and the receiving peers based on a deterministic data-to-peer assignments. PANDAS is tailor-made for Ethereum with its specific architectural assumptions and not generalizable to  other systems. In contrast, our DAS design aims to work universally as a data availability layer. 

RDA~\cite{rda_paper} is the latest development by the Ethereum Foundation as a response to the limitations of multi-hop dissemination (gossip and DHT) in adversarial settings. RDA can work for any blockchain network. To avoid multi-hop routing, RDA organizes the nodes into a grid P2P network where a node knows every other in the same row or column. Then RDA divides the data block into equally-sized chunks of symbols, assigning each chunk to a column of nodes, all of which will fully store this chunk.  The sampling of an arbitrary symbol is simple: find the destination column of the containing chunk and query every node in this column. It is assumed that each column contains at least an honest node, and so this node  will return the symbol.  RDA is provably secure against all Byzantine attacks.
However, RDA incurs substantial data duplication due to full-column replication. Because a chunk has to be broadcast to all column-wise nodes, the communication cost is also expensive. To compare, our aim is for a solution that is as provably secure as RDA but more scalable with better efficiency in both storage and communication.

\section{Preliminaries}\label{sec:prelim}
We present some background about DAS, cryptographic commitment, and random linear network coding (RLNC) which are needed for presenting our proposed solution later. 

\textbf{Notation. } A raw data block is a sequence of symbols,  $m = (m_1,\dots,m_K) \in \Gamma^{K}$,  from some alphabet $\Gamma$.  An 1D erasure code is a function, $\mathcal{C} : \Gamma^{K} \to \Lambda^{N}$, that encodes a block $m$ into an extended sequence $\hat m = (\hat m_1,\dots,\hat m_N) \in \Lambda^{N}$ from some alphabet $\Lambda$, such that any $t$ (reconstruction threshold) correct positions of $\hat m$ suffice to reconstruct the original block. For verification purposes, each symbol $\hat m_i$ at position $i \in [N]$ has a corresponding opening $\tau_i \in \Xi$ which is generated to certify that the symbol $\hat m_i$ is consistent with a public commitment. The codeword exposed to verifiers is 
\(
\pi = \bigl((\hat m_1,\tau_1),\dots,(\hat m_N,\tau_N)\bigr) \in (\Lambda \times \Xi)^{N}.
\)

\subsection{Data Availability Sampling (DAS)}
\label{subsec:das}

\subsubsection{Erasure Code Commitment (ECC)}
\label{preliminaries:ecc_scheme}

ECC is a core component of any DAS scheme. This enables a publisher to commit to erasure-coded data and later provide verifiable openings for individual positions. With this,  every symbol of the encoded sequence can be checked for consistency with a single committed codeword; this is the cryptographic foundation for sampling-based verification \cite{hasw2023}.

Formally, for an erasure code $\mathcal{C} : \Gamma^{K} \to \Lambda^{N}$, an ECC scheme is a quadruple of algorithms,
\(
\mathsf{CC} = (\mathsf{CC.Setup}, \mathsf{CC.Com}, \mathsf{CC.Open}, \mathsf{CC.Ver})
\). 
The setup algorithm generates public parameters,
\(
\mathsf{CC.Setup}(1^\lambda) \to \mathsf{ck}.
\)
To commit to a message $m \in \Gamma^{K}$, the committer executes
\(
(\mathsf{com},\mathsf{St}) \leftarrow \mathsf{CC.Com}(\mathsf{ck}, \hat m),
\)
producing a succinct commitment and internal state. For each position $i \in [N]$, an opening is computed by
\(
\tau_i \leftarrow \mathsf{CC.Open}(\mathsf{ck},\mathsf{St},i),
\)
allowing any verifier to check consistency with the committed codeword. Verification is deterministic,  given by
\(
\mathsf{CC.Ver}(\mathsf{ck},\mathsf{com},i,\hat m_i,\tau_i) \to b \in \{0,1\}.
\)

\subsubsection{Index Sampler}

The index sampler specifies which positions of the encoded sequence to be queried by independent samplers for verification purposes,   such that a sufficiently large and diverse set of fragments is collectively observed. 
Formally, an index sampler with quality function \(\nu\) is a probabilistic algorithm, \( \mathsf{Sample}(1^{Q},1^{N}) \to (i_j)_{j \in [Q]} \), that outputs \(Q\) indices in \([N]\). Its quality is characterized by the probability that \(\ell\) independent samplers collectively observe at most \(\Delta\) distinct positions: 
\[ 
\Pr\!\left[\, \bigl|\cup_{r \in [\ell]} \{\, i_{r,j} \mid j \in [Q] \,\}\bigr| \le \Delta \right]
   \le \nu(\Delta,N,Q,\ell),
\]
where each tuple \( (i_{r,j})_{j \in [Q]} \) is sampled independently.
The sampler therefore quantifies the risk that multiple participants fail to cover enough positions for reconstruction. 

\subsubsection{DAS Scheme}
\label{preliminaries:das_scheme}

A DAS scheme combines an ECC scheme with an index sampler. The goal is to ensure that, from accepting
sampling transcripts, the original message can be uniquely
reconstructed. Let \(\mathcal{C} : \Gamma^{K} \to \Lambda^{N}\) be an erasure code with reconstruction threshold \(t\) and reconstruction algorithm \(\Reconst\). Let \(\mathsf{CC} = (\mathsf{CC.Setup}, \mathsf{CC.Com}, \mathsf{CC.Open}, \mathsf{CC.Ver})\) be an erasure code commitment scheme for \(\mathcal{C}\), and let \(\mathsf{Sample}\) be an index sampler with quality function \(\nu\). The resulting DAS scheme is the tuple
\(
\mathsf{DAS}[\mathsf{CC},\mathsf{Sample}] = (\mathsf{Setup},\mathsf{Encode},\mathsf{V},\mathsf{Ext}),
\)
defined as follows.
\begin{itemize}
\item  The setup algorithm runs \(\mathsf{ck} \leftarrow \mathsf{CC.Setup}(1^\lambda)\);
\item  The encoding algorithm computes \(\hat m := \mathcal{C}(m)\), generates a commitment \((\mathsf{com},\mathsf{St}) \leftarrow \mathsf{CC.Com}(\mathsf{ck}, \hat m)\), and for each \(i \in [N]\) computes the opening \(\tau_i := \mathsf{CC.Open}(\mathsf{ck},\mathsf{St},i)\), producing \(\pi = ((\hat m_i,\tau_i))_{i \in [N]}\);
\item  A verifier samples indices \((i_j)_{j \in [Q]} \leftarrow \mathsf{Sample}(1^{Q},1^{N})\), retrieves the corresponding pairs, and accepts if \(\mathsf{CC.Ver}(\mathsf{ck},\mathsf{com},i_j,\hat m_{i_j},\tau_{i_j}) = 1\) for all \(j\);
\item Given \(\ell\) accepting transcripts, the extractor computes the union \(I\) of all queried indices; if \(|I| \ge t\), it reconstructs \(m = \Reconst((\hat m_i)_{i \in I})\), else outputs \(\bot\) (failure);
\end{itemize}

\subsubsection{DAS Security}  
A DAS scheme is secure when it satisfies three properties. 
\emph{Completeness} requires that honest sampling covers at least \(t\) distinct positions with high probability.  
\emph{Soundness} ensures that no adversary can produce accepting transcripts inconsistent with the committed data.  
\emph{Consistency} guarantees that all accepting reconstructions yield the same message.
To achieve these properties,  the sampler must provide sufficient coverage and the commitment scheme must satisfy \emph{position binding} and \emph{code binding}.
Since practical samplers with negligible coverage error exist \cite{hasw2023}, these two binding properties suffice for secure DAS.


\subsection{Tensor Codes for Secure DAS}

\begin{figure}[t]
    \centering
    \includegraphics[width=0.5\linewidth]{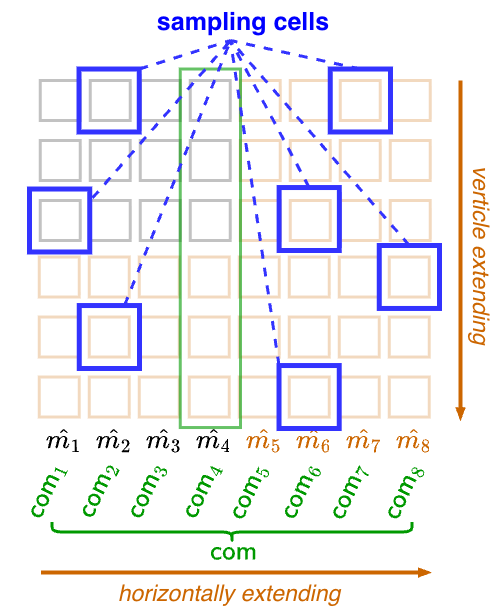}
    \caption{Tensor Codes: 2D Reed Solomon for data encoding. Sampling at the cell level. KZG commitment for each column.}
    \label{fig:tensorcode}
\end{figure}
\label{subsec:tensorcodes}

So far we have assumed an 1D erasure code. PeerDAS \cite{peerdas2024} is  the first formally specified and deployed solution for secure DAS in blockchain networks following this approach.  
Tensor Codes \cite{hasw2023} extends from PeerDAS by applying 2D Reed-Solomon erasure coding. Here, the original block and extended block are seen as $K \times K$  and  $N \times N$ matrices of symbols, respectively. The sampling is at the cell level. Each cell corresponds to a single polynomial evaluation and admits a succinct KZG opening, allowing light clients to sample arbitrary cells.
This finer granularity significantly improves sampling efficiency. A few dozen random cell samples per client already provide strong availability guarantees for practical blob sizes \cite{pandas2025}.

As illustrated in Figure~\ref{fig:tensorcode}, each encoded  $\hat{m}_i$, a column in the 2D extended block,   is interpolated into a polynomial and committed via a Zaverucha-Goldberg scheme (KZG) \cite{kate2010polycommit} commitment $\mathrm{com}_i$. 
Separate openings $\tau_i$ are computed for individual cell evaluations within the column. 
Given a randomly sampled cell (symbol) at any position and the commitment vector $(\mathrm{com}_1, \mathrm{com}_2, \dots)$, a light client can verify both its content and cell position using the corresponding opening and standard KZG verification. 
We leverage this Tensor Codes structure in our ECC scheme.

\subsection{Random Linear Network Coding (RLNC)}
\label{sec:rlnc}

RLNC is a  coding technique based on the idea that we can recover the original data given a suﬃcient number of  random linear combinations of its pieces.  RLNC-based storage systems exhibit strong robustness under node churn, since a newly joined node can generate valid coded fragments directly from existing ones without reconstructing the original data \cite{fitzek2014distributed}.

\textbf{Encoding.}
Consider a data object consisting of $L$ symbols over a finite field $\mathbb{F}_q$. We decompose it into \(k\) equally-sized pieces
\(
X = (x_1,\dots,x_k)^\top \in \mathbb{F}_q^{k \times (L/k)}.
\)
An encoded fragment, or encoded piece, is a random linear combination of these $k$ raw pieces,
\( 
c = gX = \sum_{i=1}^k g_i x_i,
\)
with coding vector
\(
\mathbf g = (g_1,\dots,g_k) \in \mathbb{F}_q^k.
\)
The size of an encoded piece equals the raw-piece size ($1/k$ of object size). 

\textbf{Decoding.}
To represent \(m\) coded pieces, we write
\(
C = GX,
\)
with coding matrix \(G \in \mathbb{F}_q^{m\times k}\). Given $C$, we can reconstruct $X = G^{-1}C$ if
\(
\mathrm{rank}(G)=k.
\)
This happens almost surely with sufficiently large $k$ and $m$ if the coding vectors are independently generated at random.

\textbf{Re-coding.}
New coded pieces can be produced from existing ones without decoding.  This is useful for message forwarding in a network. Suppose that a node receives a
coded piece \((c, \mathbf g)\), 
it  can generate a new coded piece $c'= \sum_{i=1}^k \alpha_i g_i x_i = \mathbf g' X$ with corresponding coding vector 
\( \mathbf g' = (  \alpha_1  g_1, \dots \alpha_k g_k) \)  using any random vector \((\alpha_1, \alpha_2, \dots,\alpha_k) \in \mathbb{F}_q^k\).   
This property allows new coded pieces  to be generated, replaced, or propagated using only locally available pieces. RLNC will be used in the storage and retrieval tasks of  our DAS system. 



\section{CDA: Our DAS Solution}
\label{sec:design}
We propose a new DAS design called Coded Distributed Array (CDA). This name comes from observing that DAS resembles a robust distributed array problem. The data block to distribute is a 2D array and quick sampling of  arbitrary cells is needed for availability verification. We present the details of CDA below in terms of three key components:  1) \underline{Data encoding}: how an original block is encoded and broken into encoded pieces such that the block can be verified of availability by random sampling of these pieces;
2) \underline{Commitment generation}: how commitment proofs are generated such that a cell/symbol can be verified of its correctness in terms of both content and matrix position;
and
3) \underline{Networking layer}: how the nodes in the system maintain connectivity and communicate to serve the task of data storage and sampling for verification purposes.

\subsection{Data Encoding}

\begin{figure}[t] 
\centering
\includegraphics[width=0.9\linewidth]{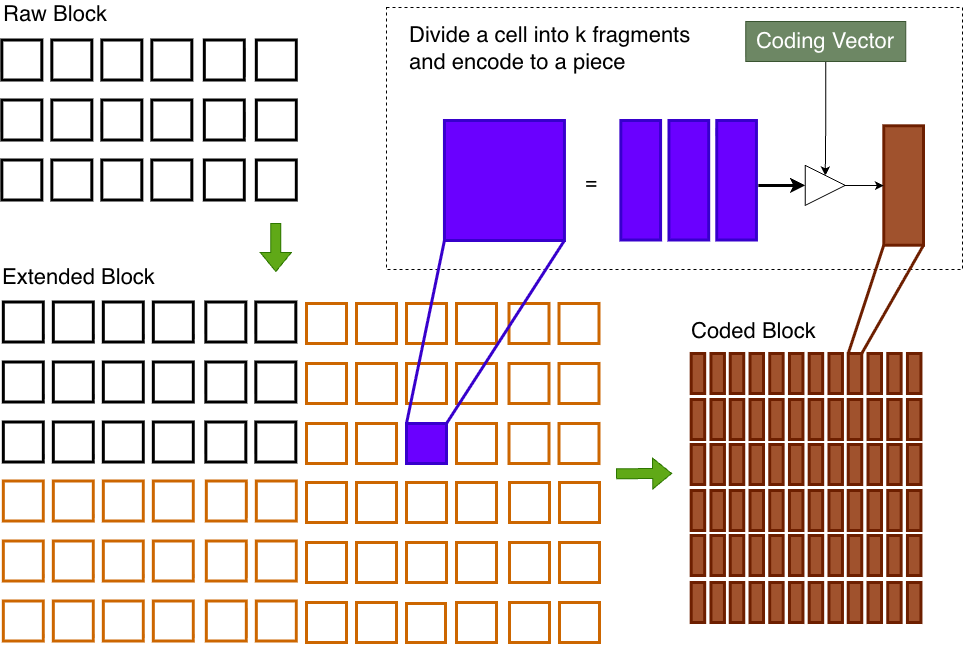}
\caption{RLNC encoding at the cell level of a data block.}
\label{fig:encoding}
\end{figure}

The process of encoding a block is illustrated in Figure \ref{fig:encoding}. Consider a data block $m$ (a matrix of $K \times K$ symbols). To produce an encoded version of this block, there are three steps.  First, 2D Reed-Solomon erasure coding is applied resulting in an extended block $\hat{m}$ (a matrix $N \times N$ of  symbols). Second, each symbol at cell $[r,c]$ is horizontally divided into $k$ equally-sized pieces:
\[ 
\hat{m}[r,c ] = (\hat{m}[r,c,1], \hat{m}[r,c,2], \dots, \hat{m}[r,c,k]).
\]
Third, a RLNC piece is generated using some random coding vector $\mathbf{g}$, 
\(  
\hat m'[r,c ] = \mathbf g \cdot \hat m[r,c] = \sum_{i=1}^k g_i \hat{m}[r,c,i].
\)
Hence we obtain a coded block $\hat{m}'$ that is a $N \times N$ matrix where each cell is the RLNC piece of the symbol at the original cell. Our key novelty is the additional RLNC layer at the cell level.

In CDA, each raw block $m$ has one extended block $\hat{m}$ and many coded versions, $\hat{m}'_1, \hat{m}'_2, \dots$, generated using different random coding vectors. The availability of $m$ is determined by successful sampling of random symbols in $\hat{m}$. Recovering such a symbol requires sampling enough RLNC pieces from $\hat{m}'_1, \hat{m}'_2, \dots$. Since each RLNC piece is only $1/k$ the size of a raw symbol, CDA reduces the sampling size by a factor of $k$ compared to conventional methods.

\subsection{Commitment Scheme}\label{sec:cda_commitment}
Since RLNC pieces are linear combinations of raw pieces, the ECC scheme must be \emph{additively homomorphic} over $\mathbb{F}_q$:
\begin{align*}
\mathrm{CC.Com}(a+b) &= \mathrm{CC.Com}(a) + \mathrm{CC.Com}(b),\\
\mathrm{CC.Com}(\alpha a) &= \alpha \cdot \mathrm{CC.Com}(a),
\end{align*}
for all $a,b,\alpha \in \mathbb{F}_q$. This lets us verify any RLNC piece directly from the same linear combination of committed coded pieces, without reconstructing the underlying symbol. We use KZG, which is additively homomorphic and supports position-wise openings.

\textbf{Symbol verifiability. } 
Verification of a symbol  requires a block commitment key $  \mathsf{ck}$, a commitment to the symbol cell, and an opening that proves the value at the queried position. Consider a symbol at matrix position $[r,c]$. Its   value  in the extended block is $X=\hat{m}[r,c] =(x_1, \dots, x_k)$ (a partitioned vector of $k$ pieces). The corresponding value in a  coded block $\hat{m}'$ (with coding vector $\mathbf g$) is coded piece $\hat{m}'[r,c]=\mathbf g X = \sum_{i=1}^k g_i x_i$.
Let each piece \(x_i\) have its own commitment 
$\mathrm{com}_i = \mathsf{CC.Com}(x_i)$,
opening $
\tau_i = \mathsf{CC.Open}(x_i)$.   
 KZG is additively homomorphic and so 
\begin{align*}
\underbrace{\mathsf{CC.Com}(\sum_{i=1}^k g_i x_i)}_{\mathrm{com}^{\ast}}
= \sum_{i=1}^k g_i \mathsf{CC.Com}(x_i)
= \sum_{i=1}^k g_i \mathrm{com}_i\\
\underbrace{\mathsf{CC.Open}  (\sum_{i=1}^k g_i x_i  )}_{\tau^{\ast}}
= \sum_{i=1}^k g_i \mathsf{CC.Open}(x_i)
= \sum_{i=1}^k g_i \tau_i.
\end{align*}
These algebraic operations preserve binding under the underlying $t$-Strong Diffie-Hellman assumption \cite{BB04}.

Consequently, a verifier can check the RLNC piece \((\mathbf g, D[r,c], \tau^{\ast})\) directly against
\(\mathrm{com}^{\ast}\) using the standard commitment verification algorithm, without reconstructing the original symbol.
Because KZG is linear and position-wise verifiable, these derived commitments and
openings preserve both properties of {code binding} and {position binding} as an ECC scheme. The security of CDA thus follows.

\begin{figure}[t] 
\centering
\includegraphics[width=0.9\linewidth]{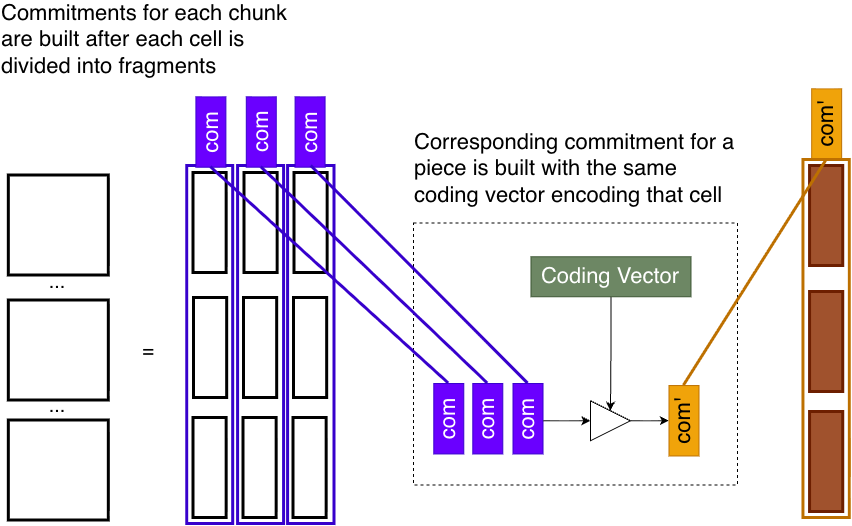}
\caption{KZG commitment computation per coded column.}
\label{fig:commitment}
\end{figure}

\textbf{Commitment computation.} We now list the steps  to compute the necessary commitments given a data block; see illustration in Figure \ref{fig:commitment}.
\begin{enumerate}
\item Extended block $\hat{m}$: because each symbol  is horizontally divided into $k$ pieces, we can think of $\hat{m}$ as a piece-matrix $\hat{m}_k$ of size $Nk \times N$  where each cell is piece. Each column in $\hat{m}$ is a $k$-supercolumn in  matrix $\hat{m}_k$ (spanning $k$ piece-columns).
\item Piece matrix $\hat{m}_k$: apply Tensor Codes to compute a KZG commitment for each column. Totally, there are $Nk$ per-column commitments.
\item Coded block $\hat{m}'$ with coding vector $\mathbf g$: each column $i$ has KZG commitment $\mathrm{com}_i$ that is the linear combination of the commitments of the piece-columns $i, i+1, ..., i+k-1$, using  the same coding vector $g$ as coefficients.
\end{enumerate}
A block's publication  contains the block commitment $\mathrm{ck}$ and $N$ KZG commitments $\mathrm{com}_1, \dots, \mathrm{com}_N$. Given this information, a coded piece under sampling can be verified to be valid.

\subsection{Network Layer}
We have established CDA's security properties and now consider the network layer. As discussed in Section \ref{sec:related}, DHT/Skip-list approaches cannot guarantee availability under malicious participants, while unstructured gossip topologies incur high dissemination and retrieval latency. RDA \cite{rda_paper} balances these extremes with a grid topology, which CDA also adopts.

\subsubsection{P2P Topology}

\begin{figure}[t]
  \centering
  \includegraphics[width=0.9\linewidth]{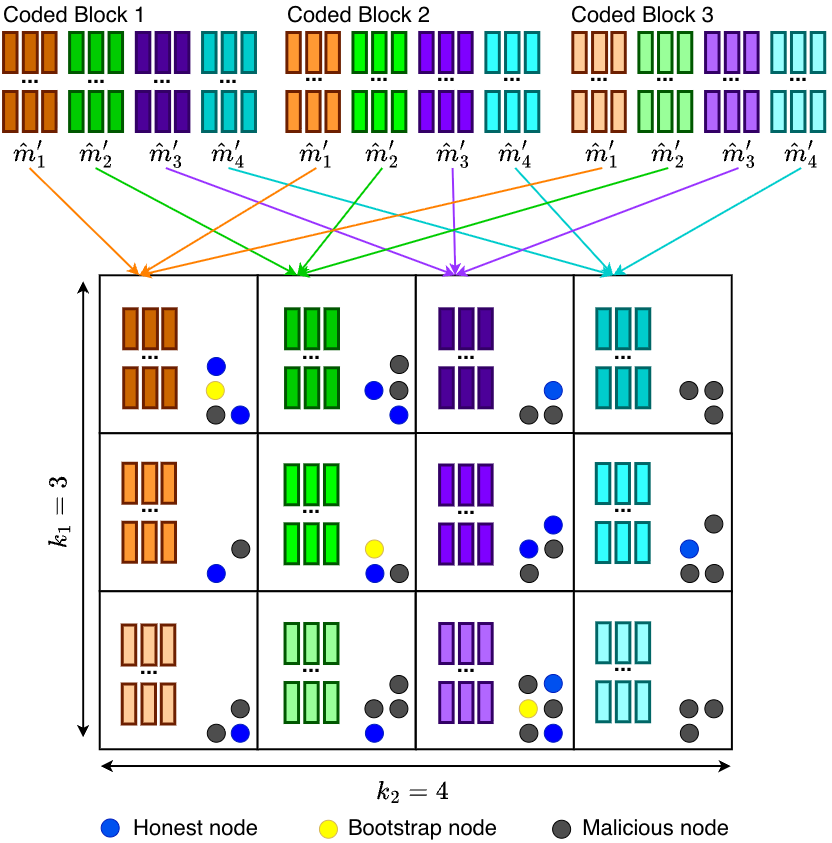}
  \caption{CDA: Assignment  of chunks to network columns.}
  \label{fig:network-topology}
\end{figure}

Define a grid of size $k_1 \times k_2$; see Figure \ref{fig:network-topology}. Nodes are assigned to random cells, called custody cells. Each node maintains contact with peers that are all the nodes in the same row or column. 
The network requires a globally-known set of bootstrap nodes such that each row has   at least one bootstrap node.  A new node joins the network and gets its peer list  by contacting   bootstrap nodes. Nodes can behave Byzantine, but there requires at least  one honest node  in every column; this is achieved when sufficient $\Omega(k_1 k_2)$ good nodes join the network. 
When a new node joins and gets its peer list, it obtains historical data
from the peers. This synchronization phase lasts for a bounded
period $\Delta_{sync}$, after which the new node is considered fully joined at the network layer.

\subsubsection{STORE Mechanism}

\begin{figure}[t] 
    \centering   \includegraphics[width=0.9\linewidth]{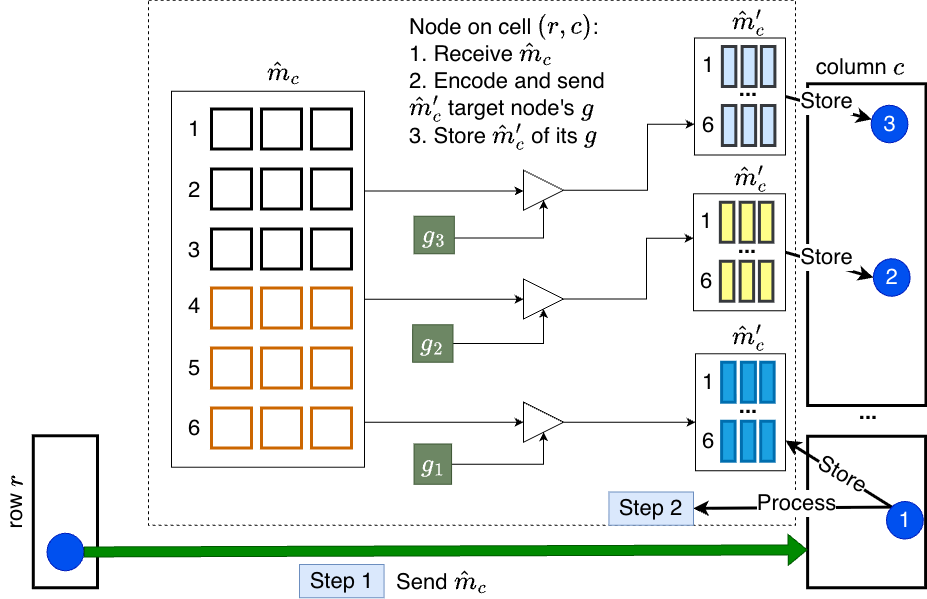} 
    \caption{STORE mechanism of each chunk $\hat m_c$: step 1) publisher node broadcasts $\hat m_c$ to peers in destination cell $[r,c]$ (node 1, ...); step 2) Each node in this cell (node 1, ...) generates a random RLNC version, $\hat m'_c$, to send to each peer in  column $c$ (node 2, 3, ...).}
    \label{fig:store}
\end{figure}

The idea is to divide the data block into $k_2$ chunks and distribute each to a corresponding network column (called the custody column, Figure \ref{fig:network-topology}).
Consider an original block $m$ with its extended $\hat m$. RLNC is applied independently to
the cells of $\hat m$, resulting in a  coded block $\hat m'$. Note that this is a $N\times N$ matrix where each cell has size of a piece. 
The dissemination is as follows; see Figure \ref{fig:store}: 
\begin{enumerate}
    \item Publisher node $P$, say on row $r$: partition extended block $\hat m$  horizontally  into $k_2$ chunks, 
\( 
\hat m = (\hat m_1, \hat m_2, \dots, \hat m_{k_2})
\). Then broadcast each chunk $\hat m_c$ to all nodes in cell $[r,c]$.
    \item Each node $Q$ in cell $[r,c]$:  independently apply RLNC on $\hat m_c$ to produce a coded chunk $\hat m'_c$ to send to each peer $R$  in column $c$. Different coding vectors are used for different nodes.  Node $Q$   also generates a coded chunk for itself to save locally. 
    \item Each node $R$ in column $c$: verify the received coded chunk and save it if valid. Duplicate chunks (having the same data-matrix position) are ignored. The verification is done per piece in the chunk, following the commitment flow  described in Section \ref{sec:cda_commitment}.
\end{enumerate}
As a result, for each original chunk $\hat m_c$, different RLNC versions, $\hat m'_c$, are stored at nodes in destination column $c$. Each node holds a version with size only $1/k$ of $\hat m_c$. Hence, the storage overhead per node in CDA is $k$ times less  than RDA.

\subsubsection{GET Mechanism} 

To verify availability of a raw block $m$, random symbols of its extended block $\hat m$ will be sampled. Consider a symbol at position $[r_1, c_1]$ of the $N\times N$ matrix of $\hat m$. This symbol is inside chunk $\hat m_c$ where 
$c = \lfloor c_1/k_2 \rfloor$. We thus need to get a sufficient number of RLNC pieces for the symbol from the nodes in custody column $c$. 
A sampling node $P$ (on row $r$) processes as follows:
\begin{enumerate}
    \item Node $P$ broadcasts the request to its peers in cell $[r,c]$. 
    \item Each node $Q$ in cell $[r,c]$: 
        \begin{enumerate}
            \item Request each peer $R$ in column $c$ for a RLNC piece (corresponding to the requested symbol).
            \item Decode the symbol from sufficient RLNC pieces. 
            \item Return the  symbol to sampling node $P$.
        \end{enumerate}
\end{enumerate}

It is possible that   the destination cell $[r,c]$  is bad without any honest node. In this case,  the sampling node $P$ will retry via bootstrap nodes to request RLNC pieces directly from nodes in column $c$. This fallback method resorting to bootstrap nodes is also employed in  RDA \cite{rda_paper}.

\begin{figure}[t] 
    \centering
\includegraphics[width=0.9\linewidth]{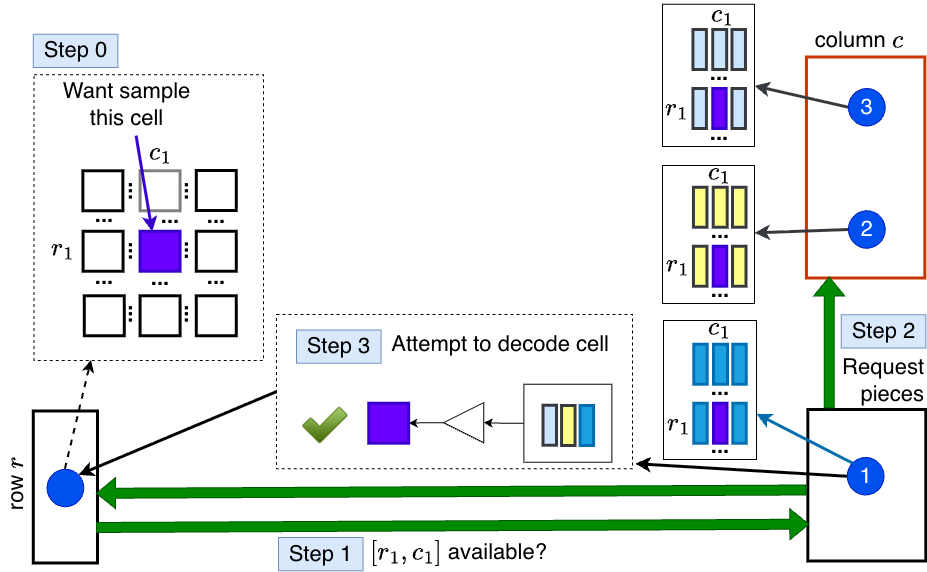} 
    \caption{GET mechanism for a symbol at data position $[r_1,c_1]$: sufficient RLNC pieces can be found in honest nodes in custody column $c$. An honest node in custody cell $[r,c]$ will retrieve and decode these pieces to reconstruct and send the symbol to the sampling node.}
    \label{fig:fullpage}
\end{figure}

The sampling success of CDA depends on the existence of sufficient RLNC pieces, which are of small size, collectively held by honest nodes in the destination column. In contrast, RDA, without RLNC coding, requires at least one big full chunk held by an honest node. Due to randomness nature of nodes joining the network, for RDA to get to the time when at least one honest node exists in each column, it is highly likely that a column also contains other honest nodes. Therefore, CDA takes advantage of this redundancy.

\section{Evaluation}
\label{sec:evaluation}
We evaluated CDA by comparing directly to RDA, the latest DAS development from Ethereum Foundation. The benchmark setting and parameter value choices follow closely the experimental setup in its published paper \cite{rda_paper}. Our evaluation source code is publicly available on Github: \url{https://github.com/Coded-Distributed-Array/benchmarks}

\subsection{Evaluation Setup}
\label{sec:eval-setting}

\textbf{Time and churn parameters.} Time is divided into synchronous rounds of 4 seconds. The system lifetime of 10 years.  
In other words, each honest must stay continuously online for $\Delta_{\mathrm{overlap}} = 6$ hours, and it gives $\Delta_{\mathrm{sync}} = 15$   minutes for a newly joined node to be synchronized into the network.

\textbf{Data parameters.} A data block is of size 
$B$ = 32MB. After 2D erasure coding, the extended block is a $256 \times 256$ matrix of size $4B = 128 \mathrm {MB}$. Each cell is thus $2\,\mathrm{KB}$,  matching the size granularity used in PeerDAS~\cite{peerdas2024}. The cryptographic proof overhead for each cell is $48\,\mathrm{B}$.
In the RLNC application in CDA,   each cell is split with fragment size $k=16$ and size of a piece $128\,\mathrm{B}$.
Each chunk of the extended block has size  
 $ 512\,\mathrm{KB}$.
The size of a coded chunk is $32\,\mathrm{KB}$. Our KZG commitment is based on the BLS12-381 curve, where a compressed G1 element occupies 48 bytes.  A  chunk contains 256 coded pieces, each accompanied by one KZG proof, hence incurring a total KZG overhead of
 $ 256 \times 48~\mathrm{B}  = 12\,\mathrm{KB}$.

\textbf{Security parameters.} We consider an adversarial regime with
\(
\varepsilon = 10\%,
\)
which bound the fraction of bad cells in each  row. The overall failure
probability over the entire system lifetime is at most $10^{-9}$.
For RDA, the network dimensions $(k_1, k_2)$ are selected to satisfy this security setting. Similarly, we can also compute the necessary $k_1, k_2$ for our CDA.

\textbf{Evaluation Metrics.}
There are four metrics: 
 1)  \underline{Commitment overhead}: The total size of commitments and cryptographic proofs, normalized by the original block size;  
2) \underline{Replication factor}: The ratio of the total amount of data stored by all honest nodes to the original block size; 
3) \underline{Propagation Cost}: The total amount of data transmitted during the publication of a block in the average case;
and 4) \underline{Synchronization cost}:
    The  amount of data a newly joining honest node must download to synchronize historical data from the custody column. 

\subsection{Security Requirement on Honest Nodes}

Since CDA and RDA share the same network layer and join-leave dynamics, we simulate both systems using exactly the same model to ensure direct comparability. As in the RDA simulation model, the network is a dynamic system of anonymous nodes  that join and leave the system according to an exogenously defined churn schedule, without modeling timing delays or network overlap. The schedule begins with a small initial set of active nodes, after which new nodes join one-by-one during a warm-up phase until the target population is reached. Once the system stabilizes, churn proceeds in steady state: in every round, a fixed number of nodes leave in FIFO order and the same number of new nodes join, so that all nodes remain active for an equal number of rounds.

\begin{figure}[t]
     \centering
     \begin{subfigure}{0.492\linewidth}
         \centering
         \includegraphics[width=\linewidth]{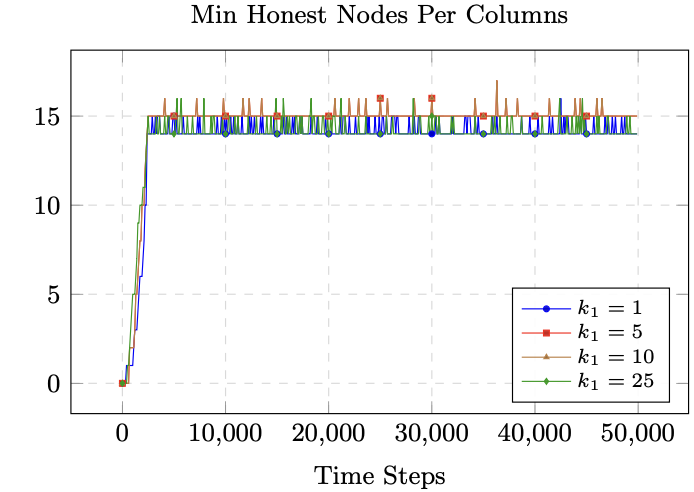}
         \caption{}
         \label{fig:min-honest-nodes}
     \end{subfigure}
     \begin{subfigure}{0.492\linewidth}
         \centering
         \includegraphics[width=\linewidth]{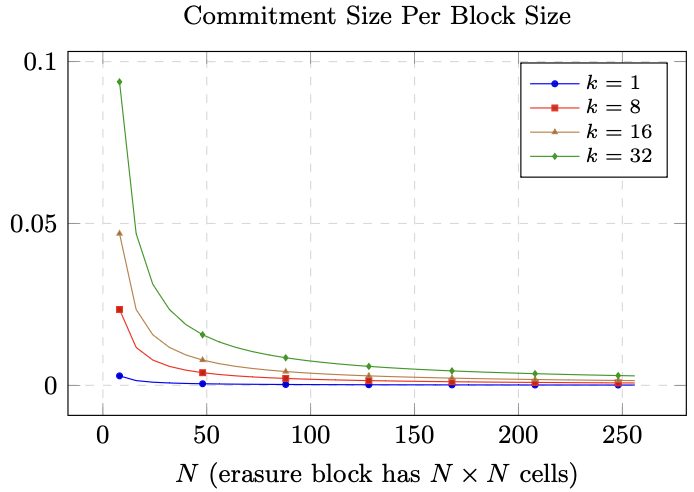}
         \caption{}
         \label{fig:commitment_overhead}
     \end{subfigure}
     \caption{Commitment size and honest-node requirements across configurations.}
\end{figure}

Figure~\ref{fig:min-honest-nodes} presents the join-leave simulation where 50 nodes join and 50 nodes leave at every step, for a total of 50,000 time steps. This setting is 10× larger than the RDA benchmark and is intended to stress-test the system and reveal any potential worst-case behavior.
The results show that each column consistently contains significantly more than one honest node; at least  12 honest nodes per column in both CDA and RDA. This behavior is stable across the entire simulation horizon. This substantiates the optimistic approach of CDA that there should be multiple honest nodes in a column  (the expected case), versus the pessimistic approach of RDA that takes advantage of only one honest node (the worst-case).

\subsection{Commitment Overhead}
Figure~\ref{fig:commitment_overhead} illustrates the ratio between commitment size and block size for different fragment sizes ($k$). The setting fragment size $k=1$ corresponds to RDA, where each cell carries a single commitment, while larger fragment sizes in CDA introduce multiple commitments per cell. Nevertheless, even for $ k \in \{8,16,32\}$, the commitment overhead is negligible and decreases rapidly with increasing block dimension, indicating that CDA’s additional commitments incur no practical cost relative to the encoded data.


\subsection{Replication Factor}

\begin{figure}[t]
     \centering
     \begin{subfigure}{0.492\linewidth}
         \centering
         \includegraphics[width=\linewidth]{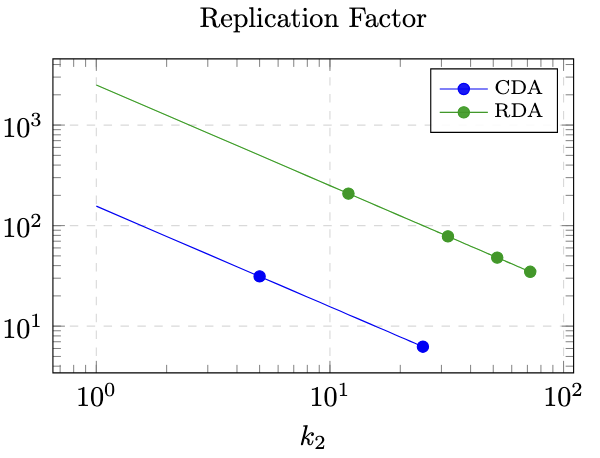}
         \caption{$M=5000$}
         \label{fig:rep_combined_5000}
     \end{subfigure}
     \begin{subfigure}{0.492\linewidth}
         \centering
         \includegraphics[width=\linewidth]{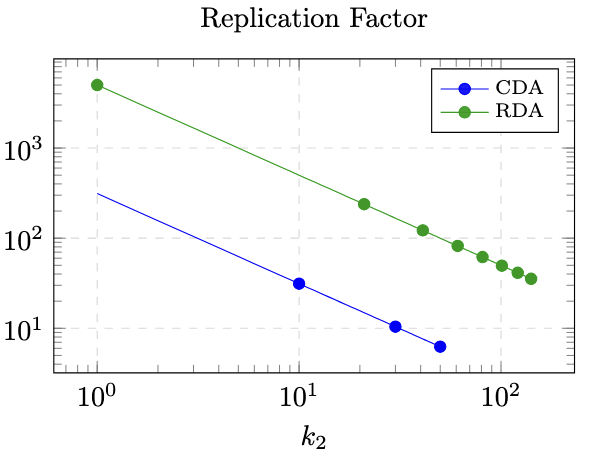}
         \caption{$M=10000$}
         \label{fig:rep_combined_10000}
     \end{subfigure}
     \caption{Replication factor: CDA vs. RDA.}
     \label{fig:rep_combined}
\end{figure}

Figure~\ref{fig:rep_combined_5000} shows the replication factor under the security constraint $M = 5{,}000$ nodes. For CDA, the feasible column dimension ranges from $k_2 = 1$ to $k_2 = 25$, corresponding to a replication factor decreasing from $156.25$  to $6.25$. In contrast, the feasible range of $k_2$ in RDA is from $1$ to $72$, hence replication factor in the range from $2{,}500$ to approximately $34.7$, a much higher range.
In practice, we will want to select the largest feasible value of $k_2$ to minimize replication. Under this choice, CDA achieves a replication factor approximately $5.5\times$ smaller than that of RDA. The same trend persists for larger network sizes, as shown in Fig.~\ref{fig:rep_combined_10000}. When the network size increases to $M = 10{,}000$ nodes, CDA continues to maintain a strictly lower replication factor than RDA, with the reduction factor approximately $5.7\times$.

\subsection{Propagation Cost}

\begin{figure}[h]
     \centering
     \begin{subfigure}{0.492\linewidth}
         \centering
         \includegraphics[width=\linewidth]{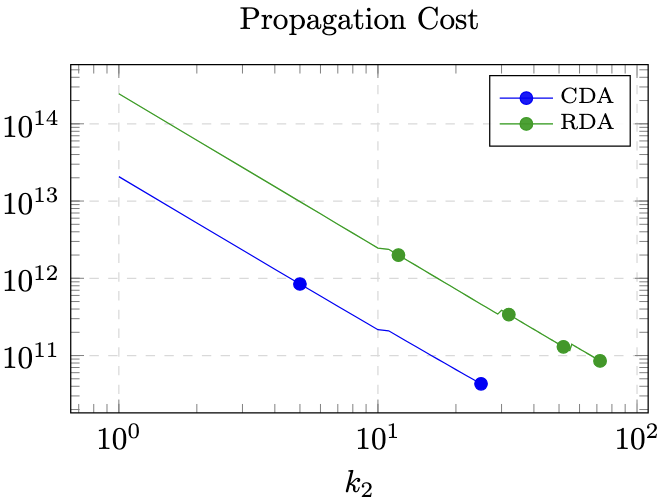}
         \caption{$M=5000$}
         \label{fig:prop_combined_5000}
     \end{subfigure}
     \begin{subfigure}{0.492\linewidth}
         \centering
         \includegraphics[width=\linewidth]{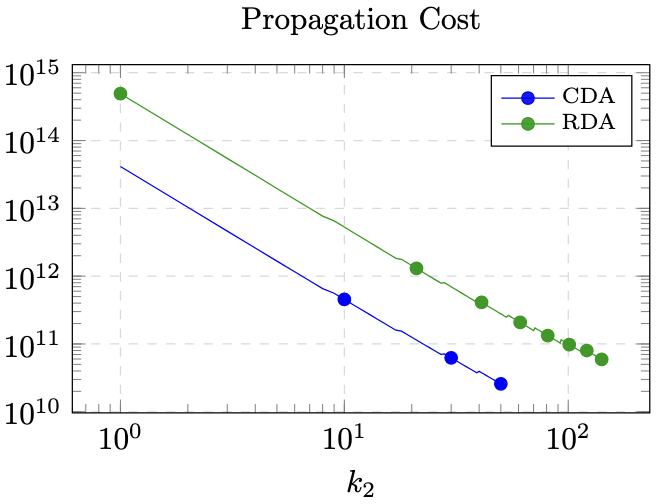}
         \caption{$M=10000$}
         \label{fig:prop_combined_10000}
     \end{subfigure}
     \caption{Propagation cost: CDA vs. RDA.}
     \label{fig:prop_combined}
\end{figure}

As shown in Fig.~\ref{fig:prop_combined_5000}, when the network has $M = 5000$ nodes, CDA achieves a propagation cost ranging from $2.07 \times 10^{13}$ byte ($19,286$ GiB) down to $4.30 \times 10^{10}$ byte ($40.05$ GiB), whereas RDA ranges from $228,200$ GiB down to $79.4$ GiB. At the optimal operating point, corresponding to the minimum propagation cost, CDA reduces the propagation cost by approximately $2\times$ compared to RDA. For larger networks, Figure~\ref{fig:prop_combined_10000} shows that when $M = 10{,}000$, the minimum propagation cost of CDA is $24$ GiB, while RDA reaches $55.1$ GiB, yielding a reduction factor of about $2.3\times$. These results demonstrate that CDA consistently incurs a lower propagation cost than RDA under identical security constraints. This is particularly important for DAS, where block broadcast efficiency directly affects network load and scalability.

\subsection{Synchronization Cost}
\begin{figure}[h]
     \centering
     \begin{subfigure}{0.492\linewidth}
         \centering
         \includegraphics[width=\linewidth]{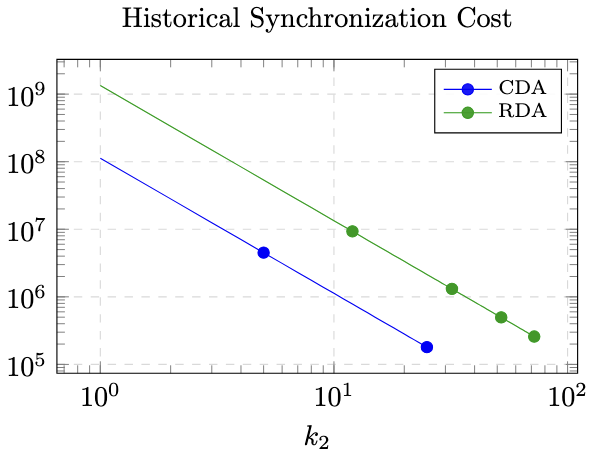}
         \caption{$M=5000$}
         \label{fig:sync_combined_5000}
     \end{subfigure}
     \begin{subfigure}{0.492\linewidth}
         \centering
         \includegraphics[width=\linewidth]{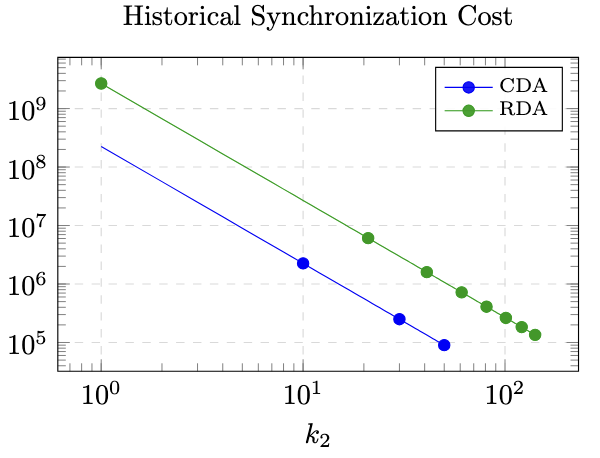}
         \caption{$M=10000$}
         \label{fig:sync_combined_10000}
     \end{subfigure}
     \caption{Synchronization cost: CDA vs. RDA.}
     \label{fig:sync_combined}
\end{figure}

As shown in Fig.~\ref{fig:sync_combined_5000}, when the network size is $M = 5{,}000$, CDA achieves a historical synchronization cost ranging from $1.13 \times 10^{8}$ byte down to $1.80 \times 10^{5}$ byte, whereas RDA ranges from $1.34 \times 10^{9}$ byte down to $2.59 \times 10^{5}$ byte. At the optimal operating point, corresponding to the minimum synchronization cost, CDA reduces the historical synchronization cost by about $1.44\times$ compared to RDA. A similar trend is observed for larger networks. When the network grows to $M = 10{,}000$ nodes (Fig.~\ref{fig:sync_combined_10000}), CDA continues to outperform RDA, achieving a minimum historical synchronization cost that is about $1.5\times$ lower. These results indicate that CDA consistently enables faster synchronization of historical state under churn. Lower synchronization cost directly improves join--leave robustness,
allowing new nodes to fetch historical data more efficiently and reducing recovery latency in practice.




\section{Conclusion}
\label{sec:conclusion}

We have proposed CDA, a novel design for DAS by jointly addressing coding, verification, and P2P networking.  By applying random linear network coding during data dissemination, CDA increases the efficiency of storage and dissemination.
The homomorphic structure of polynomial commitments enables direct verification of coded pieces without reconstructing original data. CDA is provably secure as long as  a minimal number of honest nodes joins the network. Our evaluation study has shown CDA   substantially outperform  RDA, the latest DAS development, in storage ($5\times$ better), dissemination ($2\times$), and synchronization costs ($1.4\times$) without weakening availability. A potential limitation of CDA is in the amount of coded pieces downloaded for symbol reconstruction.   This is an inevitable tradeoff because instead of replicating a full (big) chunk everywhere in the destination column, which results in only one download of the symbol from one source, CDA has to pull different pieces from different places. In the future work, we will evaluate how this tradeoff affects sampling latency and whether this effect can be negligible in experimental settings with a real blockchain network.

\bibliographystyle{IEEEtran} 
\bibliography{references} 

\end{document}